\documentclass[%
 reprint,
superscriptaddress,
amsmath,amssymb,
aps,
pra,
]{revtex4-1}

\usepackage{graphicx}
\usepackage{dcolumn}
\usepackage{bm}

\begin{document}

\preprint{APS/123-QED}

\title{Broadband space-time wave packets propagating 70~m}

\author{Basanta Bhaduri}
\author{Murat Yessenov}
\author{Danielle Reyes}
\author{Jessica Pena}

\affiliation{CREOL, The College of Optics \& Photonics, University of Central Florida, Orlando, Florida 32186, USA}

\author{Monjurul Meem}
\affiliation{Department of Electrical \& Computer Engineering, University of Utah, Salt Lake City, UT 84112, USA}

\author{Shermineh Rostami Fairchild}
\affiliation{CREOL, The College of Optics \& Photonics, University of Central Florida, Orlando, Florida 32186, USA}
\affiliation{Physics and Space Sciences, Florida Institute of Technology, Melbourne, FL 32901, USA}

\author{Rajesh Menon}
\affiliation{Department of Electrical \& Computer Engineering, University of Utah, Salt Lake City, UT 84112, USA}

\author{Martin Richardson}
\affiliation{CREOL, The College of Optics \& Photonics, University of Central Florida, Orlando, Florida 32186, USA}
\affiliation{Physics and Space Sciences, Florida Institute of Technology, Melbourne, FL 32901, USA}
\affiliation{Department of Physics, University of Central Florida, Orlando, FL 32816, USA}

\author{Ayman F. Abouraddy}
 \email{Corresponding author: raddy@creol.ucf.edu}
 \affiliation{CREOL, The College of Optics \& Photonics, University of Central Florida, Orlando, Florida 32186, USA}



\begin{abstract}
The propagation distance of a pulsed beam in free space is ultimately limited by diffraction and space-time coupling. `Space-time' (ST) wave packets are pulsed beams endowed with tight spatio-temporal spectral correlations that render them propagation-invariant. Here we explore the limits of the propagation distance for ST wave packets. Making use of a specially designed phase plate inscribed by gray-scale lithography, we synthesize a ST light sheet of width $\approx700$~$\mu$m and bandwidth $\sim20$~nm and confirm a propagation distance of $\approx70$~m.
\end{abstract}


\maketitle

`Space-time' (ST) wave packets are pulsed beams that propagate in linear materials without diffraction or dispersion \cite{Sonajalg96OL,Saari97PRL,Reivelt00JOSAA,Sheppard02JOSAA,Porras03PRE,DiTrapani03PRL,Zapata06OL,Faccio07OE,Clerici08OE,Bonaretti09OE,Turunen10PO,FigueroaBook14} by virtue of tight spatio-temporal correlations introduced into their spectra \cite{Donnelly93PRSLA,Longhi04OE,Saari04PRE}. In contrast, the behavior of the more familiar \textit{monochromatic} diffraction-free beams (e.g., Bessel, Mathieu, and Weber beams \cite{Levy16PO}) stems from their particular spatial profiles. In an ST wave packet, each spatial frequency is associated with one temporal frequency (or wavelength) \cite{Kondakci16OE,Parker16OE}, which is enforced through spatio-temporal structuring of the field; for example, via a spatial light modulator (SLM) \cite{Kondakci17NP}. Guided by this principle, we have recently synthesized a wide variety of ST wave packets, including non-accelerating Airy wave packets \cite{Kondakci18PRL}, tilted-pulse-front wave packets \cite{Kondakci19ACSP}, wave packets with group velocities tuned in free space from $30c$ to $-4c$ ($c$ is the speed of light in vacuum) \cite{Kondakci18unpub} or that travel in optical materials at $c$ \cite{Bhaduri19Optica}, incoherent diffraction-free fields \cite{Yessenov19unpub}, in addition to confirming their self-healing \cite{Kondakci18OL}.

What is the maximum distance that a ST wave packet can travel before the onset of substantial diffractive spreading? Previous realizations of ST wave packets have been limited to centimeter-scale distances. Our recent demonstration of a ST wave packet synthesized via a SLM that propagated for 6~m \cite{Bhaduri18OE} motivates investigating their ultimate propagation limit. We have recently shown theoretically that the maximum propagation distance of a ST wave packet is determined by its two internal degrees of freedom \cite{Kondakci19OL,Yessenov19Diff}: (1) the \textit{spectral uncertainty}, which is the unavoidable `fuzziness' in the association between the spatial frequencies and wavelengths; and (2) the \textit{spectral tilt angle}, which characterizes the curvature of the spatio-temporal spectral correlations and determines the wave-packet group velocity. The technical limitations of SLMs (e.g., the relatively small active area and large pixel size) pose significant obstacles to improving the spectral uncertainty. Moreover, SLMs have low energy-handling capabilities and are not available in all spectral bands. To overcome these limitations, one may utilize instead phase plates. We have used such a phase plate to synthesize broadband (bandwidth $\Delta\lambda\!\sim30$~nm) and narrowband ($\Delta\lambda\!\sim0.25$~nm) ST wave packets of spatial width $\Delta x\!\sim\!10$~$\mu$m propagating for $\sim25$~mm \cite{Kondakci18OE}.

Here we exploit a lithographically inscribed phase plate to synthesize a ST wave packet of bandwidth $\approx\!21.5$~nm that propagates for a record $\sim\!70$~m by optimizing its spatio-temporal structure. The source is a Ti:Sa-based amplified femtosecond laser system at a wavelength of $\sim\!800$~nm. Furthermore, we confirm the high laser-damage threshold of the phase plate. A theoretical model and numerical simulations provide a guide for extending this strategy to even longer propagation distances for ST wave packets, potentially on the few-kilometer-scale. Our findings pave the way for the use of ST wave packets in applications ranging from remote sensing to directed energy.

We start by considering a traditional pulsed beam $E(x,z,t)\!=\!e^{i(k_{\mathrm{o}}z-\omega_{\mathrm{o}}t)}\psi(x,z,t)$, where $\psi(x,z,t)$ is the slowly varying envelope, $\omega_{\mathrm{o}}$ is the center frequency, and $k_{\mathrm{o}}\!=\!\tfrac{\omega_{\mathrm{o}}}{c}$ is the corresponding wave number. The spatio-temporal spectrum $\tilde{\psi}(k_{x},\omega)$, which is the two-dimensional Fourier transform of $\psi(x,0,t)$, \textit{must} lie on the surface of the light-cone $k_{x}^{2}+k_{z}^{2}\!=\!(\tfrac{\omega}{c})^{2}$ \cite{Donnelly93PRSLA,Kondakci17NP}. The central requirement for synthesizing ST wave packets is to impose the spatio-temporal spectral constraint $\tfrac{\omega}{c}\!=\!k_{\mathrm{o}}\!+\!(k_{z}\!-\!k_{\mathrm{o}})\tan{\theta}$, which produces an envelope having the form
\begin{equation}\label{Eq:STWavePacket}
\psi(x,z,t)\!\!=\!\int\!\!dk_{x}\,\tilde{\psi}(k_{x})e^{i[k_{x}x-(\omega-\omega_{\mathrm{o}})(t-z/v_{\mathrm{g}})]}\!=\!\psi(x,0,t\!-\!\tfrac{z}{v_{\mathrm{g}}}),
\end{equation}
which is a wave packet undergoing rigid translation along $z$ without diffraction or dispersion at a group velocity $v_{\mathrm{g}}\!=\!c\tan{\theta}$ \cite{Yessenov18PRA}. This ideal constraint results in a dimensional reduction of the spatio-temporal spectrum $\widetilde{\psi}(k_{x},\omega)\!\rightarrow\!\widetilde{\psi}(k_{x})\delta(\omega-\omega(k_{x}))$, where $\widetilde{\psi}(k_{x})$ is the one-dimensional Fourier transform of $\psi(x,0,0)$ and $\omega(k_{x})$ is the temporal frequency associated with each spatial frequency $k_{x}$. In the spatio-temporal spectral space $(k_{x},k_{z},\tfrac{\omega}{c})$, this constraint corresponds to a spectral hyperplane $\mathcal{P}(\theta)$ that is parallel to the $k_{x}$-axis and makes an angle $\theta$ (the spectral tilt angle) with respect to the $k_{z}$-axis. Therefore, the reduced-dimensionality spatio-temporal spectrum of the ST wave packet in Eq.~\ref{Eq:STWavePacket} lies along the conic section at the intersection of the light-cone with $\mathcal{P}(\theta)$. The projection of this spatio-temporal spectrum onto the $(k_{z},\tfrac{\omega}{c})$-plane is a straight line whose slope is the group velocity along the propagation axis $\tfrac{\partial\omega}{\partial k_{z}}\!=\!c\tan{\theta}\!=\!v_{\mathrm{g}}$ \cite{Kondakci17NP,SaariPRA18,Yessenov18PRA,Kondakci18unpub,Bhaduri19Optica}.

Such ideal ST wave packets are \textit{infinite}-energy pulses. We have recently shown \cite{Yessenov19Diff} that realistic \textit{finite}-energy ST wave packets display spectral uncertainty in the association between the spatial and temporal frequencies such that the delta-function spectral correlation is replaced by $\widetilde{\psi}(k_{x},\omega)\!\rightarrow\!\widetilde{\psi}(k_{x})\widetilde{h}(\omega-\omega(k_{x}))$, where $\widetilde{h}(\omega)$ is a narrow spectral function of width $\delta\omega$ (the spectral uncertainty). The validity of this decomposition requires only that $\delta\omega$ be smaller than the bandwidth $\delta\omega\!\ll\!\Delta\omega$. It can be shown that such a ST wave packet is the product of an ideal ST wave packet traveling at a group velocity $v_{\mathrm{g}}$ and a broad uncertainty-induced `pilot envelope' of width $\tau_{\mathrm{p}}\!\sim\!1/\delta\omega$ traveling at a group velocity $c$. The maximum propagation distance $L_{\mathrm{max}}$ is thus determined by the temporal walk-off between the ideal ST wave packet and the pilot envelope \cite{Yessenov19Diff},
\begin{equation}
L_{\mathrm{max}}\sim\frac{c}{\delta\omega}\frac{1}{|1-\cot{\theta}|}.
\end{equation}
This is a surprising result: $L_{\mathrm{max}}$ is \textit{not} dictated by the beam size or the pulse width; instead, only by the internal degrees of freedom of the ST wave packet: the spectral uncertainty $\delta\omega$ and the spectral tilt angle $\theta$.

Increasing $L_{\mathrm{max}}$ thus requires reducing both $\delta\omega$ and the term $|1-\cot{\theta}|$; i.e., $\theta\!\rightarrow\!45^{\circ}$ ($v_{\mathrm{g}}\!\rightarrow\!c$). If $v_{\mathrm{g}}\!=\!(1+\Delta)c$, where $\Delta\!=\!\tan{\theta}-1\!\ll\!1$, then at $\lambda_{\mathrm{o}}\!\sim\!800$~nm we have:
\begin{equation}\label{Eq:RuleOfThumb}
L_{\mathrm{max}}\sim\frac{1}{10\cdot\delta\lambda(\mathrm{in\,\,pm})\cdot|\Delta|}\,\,\, \mathrm{m},
\end{equation}
where $\delta\lambda\!=\!\tfrac{\lambda_{\mathrm{o}}^{2}}{2\pi c}\delta\omega$ is the spectral uncertainty on the wavelength scale. For example, if $\delta\lambda\!=\!25$~pm and $\Delta\!=\!7\times10^{-4}$, as realized in \cite{Bhaduri18OE} using a diffraction grating of width 25~mm and setting $\theta\!=\!44.98^{\circ}$ via a SLM, we have $L_{\mathrm{max}}\!\sim\!6$~m. 

\begin{figure}[t!]
\centering
\includegraphics[width=8.6cm]{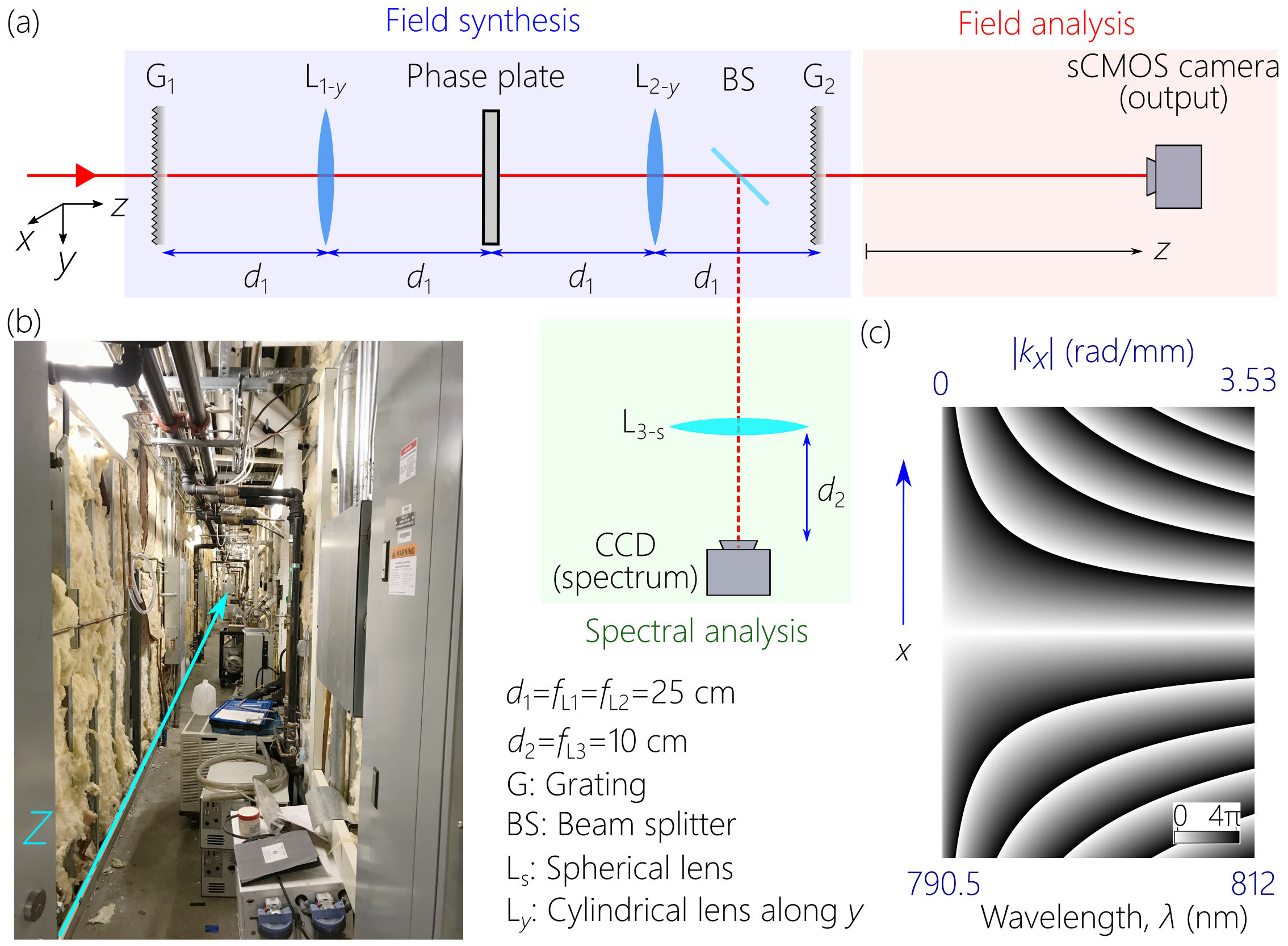}
\caption{\small{(a) Optical setup for synthesizing and characterizing ST wave packets. The setup resided in a laboratory, and the ST wave packet is directed by mirrors to a service chase along which the sCMOS camera was scanned. (b) Photograph of the service chase along which the ST wave packet was recorded. (c) The 2D phase pattern realized using the phase plate. A key is provided for the abbreviations and the distances in the setup and the focal lengths of the lenses.}}
\label{Fig:Setup}
\end{figure}

\begin{figure}[t!]
\centering
\includegraphics[width=8.6cm]{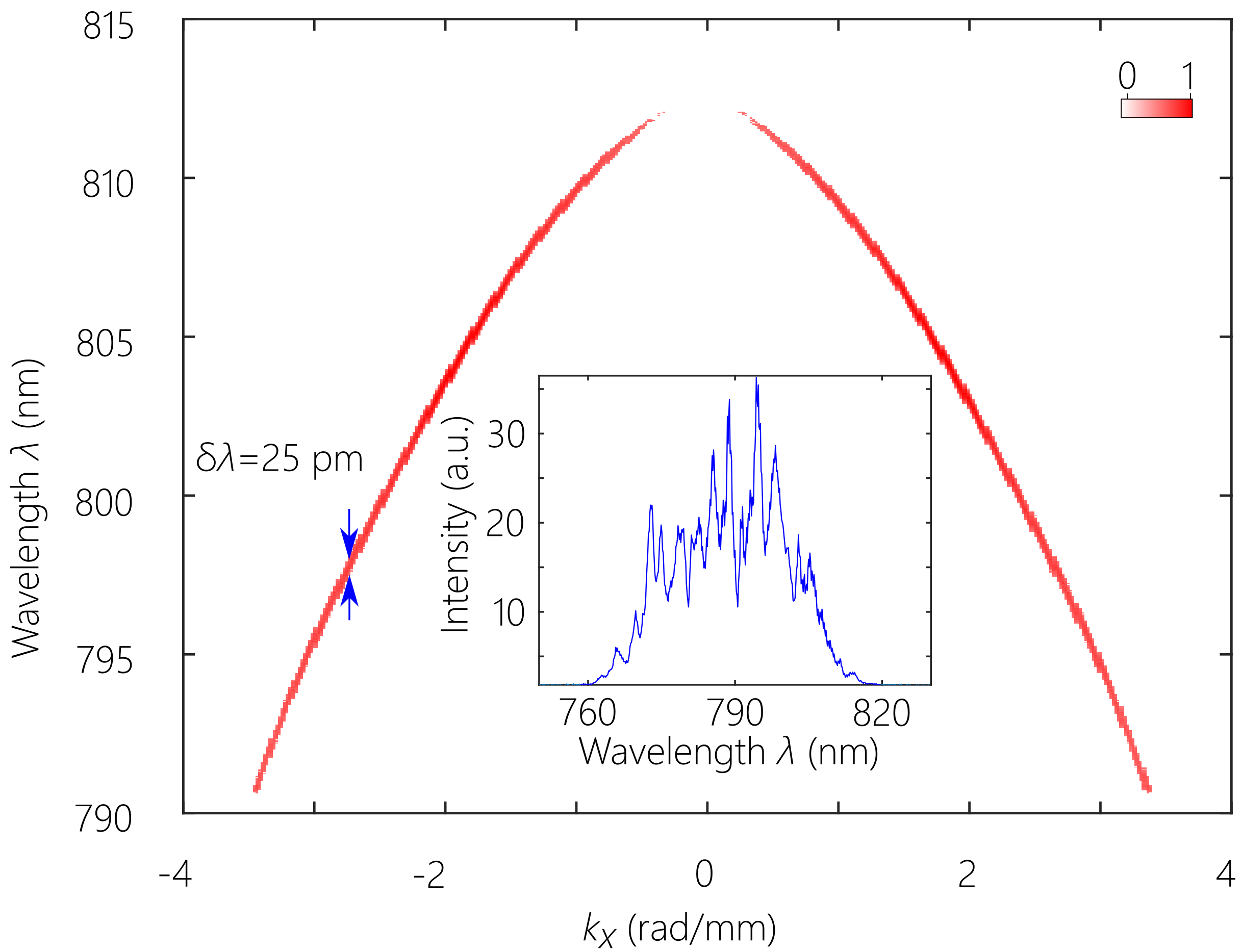}
\caption{\small{Measured spatio-temporal spectral intensity of the ST wave packet $|\widetilde{\psi}(k_{x},\lambda)|^{2}$ as recorded by the CCD camera in the setup shown in Fig.~\ref{Fig:Setup}(a); $\delta\lambda\!\approx\!25$~pm, $\Delta\lambda\!\approx\!21.5$~nm, and $\Delta k_{x}\!\approx\!3.53$~rad/mm. The inset shows the spectrum of the input (MTFL) laser.}}
\label{Fig:Spectrum}
\end{figure}

\begin{figure*}[t!]
\centering
\includegraphics[width=13cm]{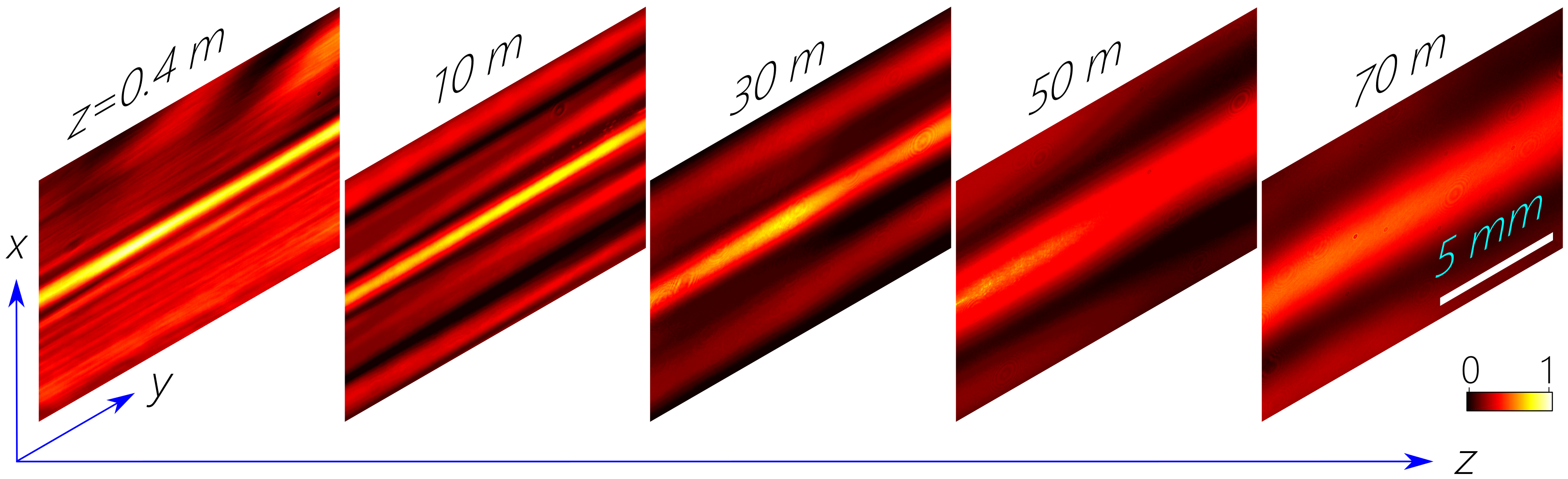}
\caption{\small{Measured time-averaged transverse intensity profiles of the ST wave packet along the propagation axis at $z\!\approx\!0.4$, $10$, $30$, $50$, and $70$~m using the sCMOS camera; see Fig.~\ref{Fig:Setup}.}}
\label{Fig:BeamProfile}
\end{figure*}

The impact of the beam size enters indirectly as follows. The temporal and spatial bandwidths of a ST wave packet, $\Delta\omega$ and $\Delta k_{x}$, respectively, are related through $\tfrac{\Delta\omega}{\omega_{\mathrm{o}}}\!=\!|f(\theta)|(\tfrac{\Delta k_{x}}{k_{\mathrm{o}}})^{2}$, where $f(\theta)\!=\!1/(\cot{\theta}-1)$ \cite{Kondakci16OE,Kondakci17NP,Yessenov18PRA}. If $\theta\!=\!\tfrac{\pi}{4}+\delta$, then $\tan{\theta}\!\approx\!\tfrac{1+\delta}{1-\delta}$ and thus $|f(\theta)|\!\approx\!\tfrac{1}{2\delta}\!\gg\!1$, where $\delta$ is in radians. Therefore, for a \textit{fixed} bandwidth $\Delta\omega$, increasing $L_{\mathrm{max}}$ (increasing $|f(\theta)|$) requires reducing $\Delta k_{x}$ and thus increasing the beam width. One may alternatively maintain the beam size but \textit{increase} the bandwidth $\Delta\omega$. Counter-intuitively, \textit{shorter} wave packets travel longer when the internal degrees of freedom $\delta\omega$ and $\theta$ are held fixed. 

In the experiment we report here, we designed a phase plate that introduces spatio-temporal correlations into the field spectrum corresponding to $\theta\!\approx\!45.00011^{\circ}$ with $\Delta\lambda\!\approx\!20$~nm. This may appear too restrictive a target for practical realization. However, $\theta$ is \textit{not} an angle in physical space, but a measure of the curvature of the spatio-temporal spectrum. As shown in \cite{Bhaduri18OE}, it is straightforward to create correlations in which $\theta$ approaches $45^{\circ}$. The phase plate utilized was fabricated using a gray-scale lithography technique \cite{Mckenna10Conf}, whereupon positive-tone photoresist (S1813, MicroChem) \cite{Shipley1813} was spin-coated on a RCA-cleaned, double-side polished 2-inch-diameter soda lime glass wafer at 1000~rpm for 60~s followed by baking in an oven at $110^{\circ}$C for 30~min. A laser-pattern-generator (Heidelberg Instruments) \cite{Heidelberg} was used to write the design. The exposed sample was then developed in AZ $1:1$ solution \cite{AZ} for 35~s followed by DI water rinse. A calibration step was performed on a different sample (prepared with the same process conditions), before writing the design to determine the depth of the photoresist at a particular gray level after development. The details of the calibration process are discussed elsewhere \cite{Mohammad17SR,Mohammad18SR}. The phase distribution realized by this plate is plotted in Fig.~\ref{Fig:Setup}(c). 

The femtosecond laser source utilized here is the Multi-Terawatt Femtosecond Laser (MTFL) \cite{Webb14CLEO,Webb16Thesis} housed at the University of Central Florida. This Ti:Sa-based chirped pulse amplification system consists of 3 stages of amplification and is capable of delivering 500-mJ-energy pulses at a wavelength of $\lambda_{\mathrm{o}}\!=\!790$~nm and spectral bandwidth of $\sim\!43$~nm, corresponding to pulses having a width of $\sim50$~fs (FWHM) at a repetition rate of 10 Hz. The beam used in this experiment was extracted directly from the regenerative amplifier (first stage of amplification) at an energy of only 1~mJ, which was further attenuated to avoid saturating the detectors.

The experimental set-up for synthesizing and characterizing ST wave packets is shown in Fig.~\ref{Fig:Setup}(a). The MTFL laser is first expanded spatially to a diameter of $\sim\!30$~mm and is incident at an angle of $51^{\circ}$ on a reflective diffraction grating G$_{1}$ (1800~lines/mm, $65\times65$~mm$^{2}$) that disperses the femtosecond pulses along the $y$-direction. The first diffraction order is collimated and directed through a cylindrical lens L$_{1-y}$ (oriented along the $y$-direction in a $2f$ configuration) to the phase-plate that introduces the requisite spatio-temporal spectral correlations between $k_{x}$ and $\lambda$. Each column of the phase plate modulates a particular $\lambda$ by imparting a linearly varying phase along the direction orthogonal to the $y$-direction and whose slope corresponds to the value of the associated $k_{x}$. The transmitted wave front is reconstituted into a pulse by a symmetrically placed lens L$_{2-y}$ and grating G$_{2}$ (1800~lines/mm, $130\times110$~mm$^{2}$) to produce the ST light sheet. All the lenses and mirrors had dimensions of at least 50~mm, except for the phase plate whose dimensions were $25\times25$~mm$^{2}$. 

\begin{figure}[t!]
\centering
\includegraphics[width=7.4cm]{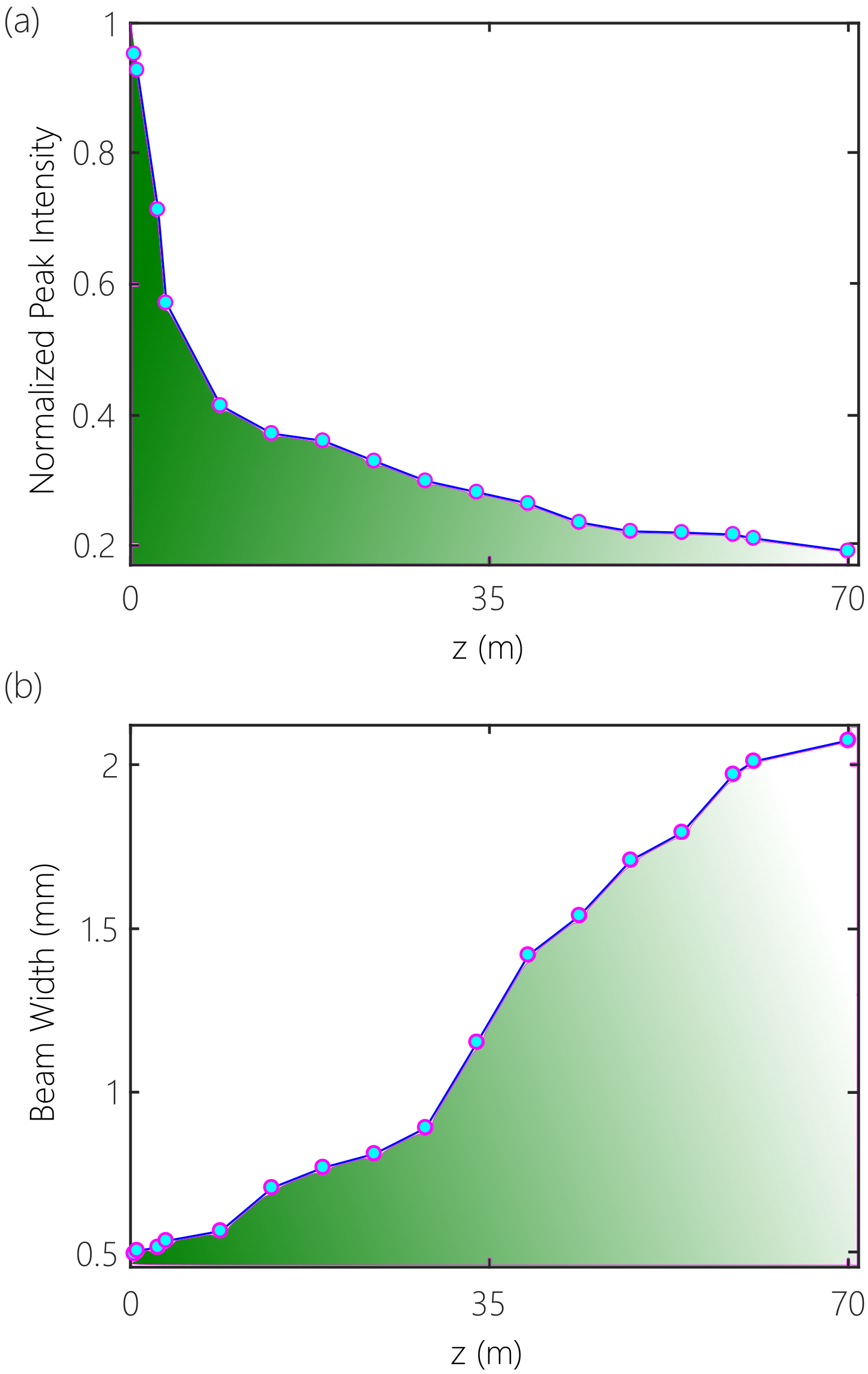}
\caption{\small{(a) Measured normalized peak of the ST wave packet profile [Fig.~\ref{Fig:BeamProfile}] along the propagation axis $z$. (b) Measured transverse width $\Delta x$ along the propagation axis $z$.}}
\label{Fig:AxialData}
\end{figure}

\begin{figure}[t!]
\centering
\includegraphics[width=8.6cm]{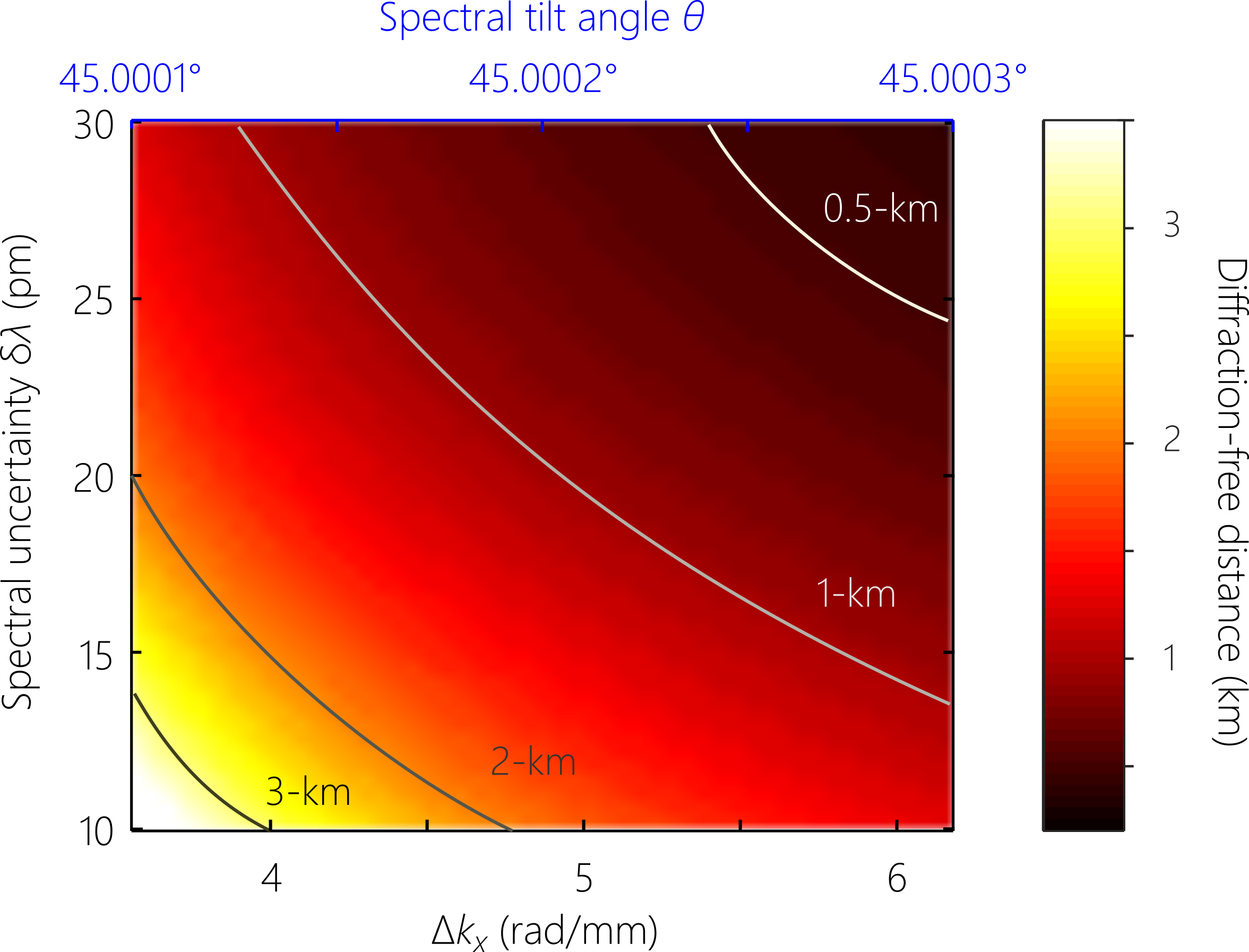}
\caption{\small{Simulations of the propagation distance for ST wave packets parametrized by the spectral uncertainty $\delta\lambda$ and the spectral tilt angle $\theta$. By fixing the bandwidth $\Delta\lambda\!=\!21.5$~nm, we can rescale $\theta$ as a spatial bandwidth $\Delta k_{x}$ (and thus beam size $\Delta x\!\sim\!1/\Delta k_{x}$).}}
\label{Fig:Theory}
\end{figure}

The ST wave packet is then directed by mirrors out of the laboratory to follow a straight path down a service chase that cuts across the CREOL building for observation [Fig.~\ref{Fig:Setup}(b)]. A sCMOS camera (Andor, Zyla) is scanned along the $z$-axis to record the time-averaged intensity of the ST light sheet every 5~m, and a CCD camera (The Imaging Source, DMK 72AUC02) records the spatio-temporal spectrum after performing a spatial Fourier transform via a spherical lens L$_{3-\mathrm{s}}$ \cite{Kondakci17NP,Kondakci18PRL,Kondakci18OE,Bhaduri18OE}; see Fig.~\ref{Fig:Spectrum}. Images of the time-averaged beam intensity in the transverse plane at different distances along the propagation direction are shown in Fig.~\ref{Fig:BeamProfile} (exposure time of each is 25~msec) and the axial evolution of the beam width and the peak intensity at the beam center $I(0,z)$ are plotted in Fig.~\ref{Fig:AxialData}.

The Rayleigh range $z_{\mathrm{R}}$ of a Gaussian beam of width 700~$\mu$m at a wavelength of 800~nm is $\approx\!48$~cm, such that $L_{\mathrm{max}}/z_{\mathrm{R}}\!\approx\!145$ here. However, it can be shown that the ratio $L_{\mathrm{max}}/z_{\mathrm{R}}$ is on the order of $\tfrac{\Delta\lambda}{\delta\lambda}$ (the ratio of the full bandwidth to the spectral uncertainty) \cite{Yessenov19Diff}, which in our experiment is $\sim\!1000$. We therefore predict that the same configuration can produce ST wave packets that may propagate for $\sim\!1$~km. In moving forward, we have found that spatial filtering of the input laser or increasing the initial beam size at the diffraction grating is critical. Secondly, utilizing a 50-mm-wide phase plate is critical for producing beams propagating larger distances. We provide in Fig.~\ref{Fig:Theory} a numerical calculation of the propagation distance $L_{\mathrm{max}}$ for ST wave packets while varying $\delta\lambda$ and $\theta$. The results are in accordance with the rule of thumb in Eq.~\ref{Eq:RuleOfThumb} and show clearly that reducing the spectral uncertainty to $\sim\!10$~pm and reaching a spectral tilt angle of $\theta\!=\!45.0001^{o}$ enables us to reach $L_{\mathrm{max}}\!\sim\!3$~km with a beam width of $\sim\!1.5$~mm and bandwidth $\Delta\lambda\!\sim\!21$~nm. 

The phase plate laser-damage threshold was assessed by slowly increasing the incident fluence until instantaneous white-light damage was observed. The pulse energy of the MTFL output beam was adjusted to 10~mJ and, and the 25.4-mm-diameter beam was directed to the phase plate via a lens having $f\!=\!3$~m. The lens is initially placed in proximity to the phase plate and then gradually moved further away until damage was observed. From these parameters we calculate the beam size at the phase plate and thence the fluence damage threshold, which was thus determined to be 584~mJ/cm$^{2}$. 

The ST wave packets presented here are in the form of a light sheet with the field uniform along one transverse dimension as a result of correlating the wavelengths with the spatial frequencies along $x$ alone. Future work will focus on extending our approach to both transverse coordinates and thus producing wave packets in the form of a needle with higher local intensities. Because ST wave packets are an example of extending the concept of `classical entanglement' from discretized degrees of freedom (e.g., polarization and spatial modes) \cite{Qian11OL,Kagalwala13NP} to continuous degrees of freedom (time and space) \cite{Kondakci19OL}, localizing the field in both transverse coordinates along with time represents an extension to classical entanglement between three continuous degrees of freedom.

In conclusion, we have synthesized a ST wave packet of bandwidth $\approx\!21.5$~nm in the form of a light sheet of width $\sim\!700$~$\mu$m that propagates in free space for a distance of $\sim70$~m. The requisite spatio-temporal correlations are introduced into the field spectrum via a specially designed phase plate fabricated via gray-scale lithography. We have presented theoretical estimates and numerical calculations that indicate the potential for extending the propagation distance to the few-kilometer range with improvements to the current experiment.

\section*{Funding}
U.S. Office of Naval Research (ONR) (N00014-17-1-2458); U.S. Office of Naval Research (ONR) (N66001-10-1-4065); U.S. Army Research Office (ARO) MURI program on Light Filamentation Science (W911NF1110297); HEL JTO (FA9550-11-1-0001).


\bibliography{diffraction}


\end{document}